\documentclass[conference,10pt,a4paper]{IEEEtran}
\usepackage{url,amsmath,cite,caption}
\ifx\pdfoutput\undefined
  \usepackage[pdftex]{graphicx}
\else
  \usepackage{graphicx}
\fi

\begin{document}

\title{The Correlation between Structural Capital and Innovation in Indonesian Manufacturing Industry}

\author{\authorblockN{R. Wijayanti, N.G. Berliana, I.M. Nadhiroh,\\
E. Aminullah, Kusnandar, T. Fizzanty,\\
T. Handayani, R. Rahmaida, N. Laili, Kusbiantono}
\authorblockA{Research Center for Development of Science and Technology\\
Indonesian Institute of Sciences\\
Jl. Gatot Subroto 10, Jakarta 12750, Indonesia\\
Email: rini.wijayanti@lipi.go.id}
\and
\authorblockN{L.T. Handoko}
\authorblockA{Group for Theoretical and Computational Physics  \\
Research Center for Physics\\
Indonesian Institute of Sciences\\
Kawasan Puspiptek Serpong, Tangerang 15310, Indonesia\\
Email: handoko@teori.fisika.lipi.go.id}
}


\maketitle

\begin{abstract}
The statistical relationship between structural capital and innovation in Indonesian manufacturing industries is presented. The correlation is constructed using recent survey data on the contribution of structural capital to the innovation processes in the industries. The correlation is represented quantitatively using the recently developed Intellectual Capital and Innovation (ICI) index involving all components of intellectual capital and its role to enable innovation in a manufacturing industry. However, the paper is focused only on the contribution of structural capital component. Using the available data it is shown that the correlation is highly depending on the scale and characteristics of each manufacture. It is also argued that the ICI index is able to quantitatively prove the dominant components in innovation processes for each class of manufacturing industries.
\end{abstract}


\section{Introduction}
\label{sec:intro}

\PARstart{I}{ncreasing} global competition in recent decades requires companies around the globe to continuously innovate in order to improve its competitiveness. In such environment, most of companies depend on neither natural and capital resources, nor the so  called tangible assets. On the other hand, they likely utilize the intangible assets as an intellectual capital and certain knowledges related to long-life experiences and particular skills. The most important intellectual capital is the human resource (HR)-based competent knowledge (knowledge-based workers) who have acquired multi skills (multi-skill worker). It is widely believed that the intellectual capital has a great potential to enhance the company competitiveness for long run. Such intellectual capital could be in encouraging creativities in improving productivity, realizing timely and efficient production process, and so on \cite{acs,castro,edquist,grif}. These phenomena affect not only the global companies, but also the local players in certain regions \cite{myt}.

Therefore, the competitive advantage of companies are nowadays assessed from its ability to innovate. Some previous studies indicate that there is a significant relationship between innovation and competitiveness. For example, using Solow Growth model it has shown that most of the increased output per capita in the United States in period of 1909-1949 was dominantly caused by the technology development \cite{solow}. The model is actually aligned with the current conditions in developed countries as Japan and Western Europe. Innovation itself is an implementation of something (can be some product, method, organization, etc.) in an entirely new or significantly different \cite{oecd1,oecd2}. In other words, innovation is the result of the development of ideas and knowledge which lead to certain economic value.

Furthermore, intellectual capital is the result of three main components of organization relating to knowledge and technology which could generate added values to the company. Those components are classified as human, structural and relational capitals \cite{leif,maddock}. Human capital is the main source of knowledges, skills and competencies of the organization. Therefore, human capital reflects the company's collective ability to produce innovation. Moreover, structural capital is the company's ability to realize optimal business and intellectual performances. So, by definition it is obvious that there is a strong relationship between intellectual capital and innovation capability of a company.

The study of intellectual capital in  Indonesian has not been intensively done. In particular, it is not trivial to investigate whether Indonesian companies are utilizing intellectual capital in developing its competitiveness or not yet. There is a common perspective that Indonesian companies tend to yet use the conventional ways in building their business and competitiveness. The company still depend much on natural resources and cheap labor. Not surprisingly, the majority of products produced by manufacturing companies Indonesia has low technological content as found in the Indonesia Science and Technology Indicators in 2000 \cite{mei}. Though, in order to enhance the competitiveness,  especially with the trade tariff exemption agreement between countries such as ASEAN China Free Trade Area (ACFTA), Indonesian companies should improve themselves to have more competitive advantage based on creative innovations.
In this paper, the contribution of intellectual capital, in particular the structural capital, on innovation performance in Indonesian manufacturing companies is studied. The detail analysis is classified according to the technology intensity, the labor scale and also the capital scale of companies. The quantitative analysis is performed using recently developed  Intellectual Capital and Innovation (ICI) index \cite{rini}. The index incorporates all components of intellectual capital and its role to enable innovation in a manufacturing industry. It is shown that the index is able to extract core information related to the correlation between structural capital and innovation performance. All analysis are using previous data obtained from the survey on Research and Development in the Indonesia Manufacture Industries conducted by the  Research Center for Development of Science and Technology (Pappiptek) LIPI in 2009 \cite{mei}.

The paper is organized as follows. After this introduction, a short review on the ICI index is given in Sec. \ref{sec:metoda}. The analysis using 2009 data is performed in Sec. \ref{sec:analisa} and the paper is ended with a summary and conclusion.

\section{Methodology}
\label{sec:metoda}

Now let us briefly review the ICI index used to analyze the data in the paper \cite{rini}. The index is intended to visualize the binary data obtained from any surveys to overcome the problem of 'missing information'. The problem frequently occurs when one deals with binary data. Binary data is quite handy and easy to process, while it tends to provide too rough and in some sense difficult to interpret since the final results are within the median values. This problem arises due to extensive usage of averaging the value improperly. 
Through ICI index, provides alternative insight by enabling qualitative interpretation of quantitative results through images obtained from visualizing the data. 

\subsection{ICI index}
 
The index is based on few assumptions. First, each component contributing to the main observables can be ordered according to its significances. Each component is then assigned with certain values accordingly, started from the maximum coefficient ($C_M$) with a universal decreasing factor ($C_D$). These yield a set of coefficients,
\begin{eqnarray}
  C_i & = & \left\{ C_M, C_D \times C_M, C_D^2 \times C_M, \cdots, 
  \right. \nonumber \\
  && \left. C_D^{n-1} \times
C_M, C_D^n \times C_M, \cdots \right\} \; ,
  \label{eq:ci}
\end{eqnarray}
where $m$ denotes the total number of components and the minimum value of $C_i$ should greater than null. It should be noted that the value of $C_D$  is arbitrary and has no meaning since it would generate the relative values among components.

Further, the index is given by the master equation, 
\begin{equation}
  I_\mathrm{X} = \frac{1}{C_T} \sum_{i=1}^n \left[ \left( C_i \times V_i
\right) + f \left( 1 - N_i \right) \right] \; ,
  \label{eq:i}
\end{equation}
for n number component with X is whatever the index name, that is the Innovation Performance (IP) and the Intellectual Capital (IC) in the present paper. $V_i = (0, 1)$ is the binary data, while $N_i = {1,2, …, n}$ represents the difference for components with same significances. $f$ is just a universal multiplication factor. $C_T$ is the sum of all component coefficients to make the index in Eq. (\ref{eq:i}) is dimensionless and can be represented further in percentage (\%).

The parameter $N_i$ has been introduced to distinguish subsequent components with same relevances. Consequently, for single component without overlapping relevance with its neighborhood has $N_i = 1$. On the other hand, the parameter $f$ is intended to adjust the decreasing scale between two neighboring components. Mathematically it is constrained by, 
\begin{equation}
  0 < f < C_D^{n-1} \times C_M \frac{ 1 - C_D }{N_M - 1}  \; ,
  \label{eq:f}
\end{equation}
Here, $N_M$ represents the maximum number of $N_i$ and its value is always greater than 1. 

\subsection{The data of manufacture industry}

In this study, the Indonesian manufacture industry is classified into several categories, that is,
\begin{enumerate}
\item Based on technology intensity :\\
Following the OECD standard \cite{oecd2}, it consists of high-technology, medium high-technology, medium low-technology and low-technology manufactures. However, for the sake of simplicity in the study adopts two classifications : high-technology and low-technology.
\item Based on business scale :\\
In Indonesia, the company scale is determined by the labor scale according to the Statistical Center Bureau and the capital scale according to the Ministry of Trade. The present study adopts the labor scale and capital scale. For labor scale there two classes : small (below 50 employees) and large (above 50 employees), while for capital scale : low (below Rp. 500 M), medium (between Rp. 500 M and Rp. 50 B) and large (above Rp. 50 B).
\end{enumerate}

The survey was conducted in 2009 against 1000 manufacture companies in Indonesia \cite{mei}. The innovation performance indicator comprises of innovations on goods (IP-1), services (IP-2), product development prior to the competitors (IP-3), production processes (IP-4), procuring processes (IP-5), supporting processes (IP-6) and any processes prior to the competitors (IP-7). Meanwhile, the structural capital covers internal R\&D (SC-1), R\&D of holding company (SC-2), external R\&D infrastructure acquisition (SC-3), external knowledge acquisition (SC-4), training (SC-5), patent (SC-6), trademark (SC-7), copyright (SC-8), CA (SC-9), trade secret (SC-10), design (SC-11) and pioneering work (SC-12). All components above are in the binary form, i.e. yes or no. 

\section{Data analysis}
\label{sec:analisa}

\begin{table}[t!]
\caption{The assigned order ($i$) of IP's components and its respective values of $N_i$ and $C_i$ for manufactures with large and small labor scales.}
\label{tab:nisdm}
\begin{tabular}{|l|c|c|c|c|c|c|}
\hline
  & \multicolumn{3}{|c|}{large labor scale}  & \multicolumn{3}{|c|}{small labor scale} \\
\cline{2-7}
\multicolumn{1}{|c|}{\raisebox{1.5ex}[0pt]{IP}}  & $i$ & $N_i$ & $C_i$ & $i$ & $N_i$ & $C_i$ \\
\hline
IP-4 & 1 & 1 & 100 &2 & 1& 90\\
IP-1 & 2 & 1 & 90 & 1& 1& 100\\
IP-6 & 3 & 1 & 81 &4 & 1& 72\\
IP-5 & 4 & 1 & 72 & 3& 1& 81\\
IP-3 & 5 & 1 & 64 &5 & 1& 64\\
IP-7 & 6 & 1 & 57 & 6 & 1 & 57\\
IP-2 & 7 & 1 & 51 & 7 & 1 & 51 \\
\hline
\end{tabular}
\end{table}

\begin{table}[h!]
\caption{The assigned order ($i$) of IP's components and its respective values of $N_i$ and $C_i$ for manufactures with high- and low-intensity technologies.}
\label{tab:nitekno}
\begin{tabular}{|l|c|c|c|c|c|c|}
\hline
  & \multicolumn{3}{|c|}{high-intensity}  & \multicolumn{3}{|c|}{low-intensity} \\
\cline{2-7}
\multicolumn{1}{|c|}{\raisebox{1.5ex}[0pt]{IP}}  & $i$ & $N_i$ & $C_i$ & $i$ & $N_i$ & $C_i$ \\
\hline
IP-4 & 1 & 1 & 100 &2 & 1& 90\\
IP-6 & 2 & 1 & 90 & 4 & 1 & 72 \\
IP-1 & 3 & 1 & 81 &1 & 1 & 100\\
IP-3 & 4 & 1 & 72 & 6 & 1 & 57 \\
IP-5 & 5 & 1 & 64 &3 & 1 & 81\\
IP-7 & 6 & 1 & 57 & 5 & 1 & 64\\
IP-2 & 7 & 1 & 51 & 7 & 1 & 51 \\
\hline
\end{tabular}
\end{table}

\begin{table}[b!]
\caption{The assigned order ($i$) of IP's components and its respective values of $N_i$ and $C_i$ for manufactures with large, medium and low capital scales.}
\label{tab:nimodal}
\begin{tabular}{|l|c|c|c|c|c|c|c|c|c|}
\hline
  & \multicolumn{3}{|c|}{large scale}  & \multicolumn{3}{|c|}{medium scale} & \multicolumn{3}{|c|}{low scale} \\
\cline{2-10}
\multicolumn{1}{|c|}{\raisebox{1.5ex}[0pt]{IP}}  & $i$ & $N_i$ & $C_i$ & $i$ & $N_i$ & $C_i$ & $i$ & $N_i$ & $C_i$ \\
\hline
IP-1 & 1 & 1 & 100 &2 & 1& 90 & 2 & 2 & 90\\
IP-4 & 2 & 1 & 90 & 4 & 1 & 72  & 2 & 1 & 90\\
IP-5 & 3 & 1 & 81 &4 & 1 & 72 & 3 & 1 & 81\\
IP-6 & 4 & 1 & 72 & 3 & 1 & 81 & 1 & 1 & 100\\
IP-7 & 5 & 1 & 64 &3 & 1 & 81& 5 & 1 & 64 \\
IP-3 & 6 & 1 & 57 & 5 & 1 & 64 & 4 & 1 & 72\\
IP-2 & 7 & 1 & 51 & 7 & 1 & 51 & 6 & 1 & 57  \\
\hline
\end{tabular}
\end{table}

Throughout the analysis, let us use the values of $C_M = 100$, $C_D = 90$\% and $f = 2$ in Eqs. (\ref{eq:ci}) and (\ref{eq:i}). These values are taken in such away the constraint in Eq. (\ref{eq:f}) is satisfied. Using these parameters and Eq. (\ref{eq:ci}), one can obtain straightforwardly a set of coefficients,
\begin{equation}
  C_i  = \{100, 90, 81, 72, 64, 57, 51, 45, 40, 36, 32, 28, 25\} \; .
  \label{eq:cid}
\end{equation}

Furthermore, one should determine the level of significance of each component mentioned in Sec. \ref{sec:metoda}. The list should be determined for each class of manufactures under consideration. Therefore, there are several lists according to the classification, for instance in the present case one has two classes of technology intensity, two classes of labor scale and three classes of capital scale. Those lists are provided in Tabs. \ref{tab:nisdm}-\ref{tab:nimodal}. From the tables, one could distinguish the differences among classes of manufactures defined in the previous section. Those differences arise from different order of importances represented by $i$ and $N_i$ as well. The orders determine the value of $C_i$ subsequently. The role of $N_i$ is well illustrated in Tab. \ref{tab:nimodal} for IP-1 and IP-4 which have comparable importances, and therefore same value of $C_i$.  Though in some cases the differences between neighboringcomponents seem small, it is already enough to make distinction as proven in the results below.

The components of IC, or in the present case is structural capital (SC), are represented by namely SC-1 to SC-12 mentioned in the preceeding section. Those should be treated in the same manner as IP-1 to IP-7 in Tabs. \ref{tab:nisdm}-\ref{tab:nimodal}. Unfortuntely, its rather lengthy tables can not be given in the paper because of page limitations.

Substituting all parameters and acquired survey data for each class one could get the desired index for IP and IC respectively. However, as mentioned before one should unfortunately not expect surprising result by averaging the index. According to our study on the same data, there was no much differences among different classes of objects. Then the analysis should be described in more intuitive way by borrowing simple image processing techniques. Rather than presenting the average values as standard statistic, the obtained values for each company are represented as the distribution of point of interest.

\begin{figure}[t!]
 \centering
 \includegraphics[width=85mm]{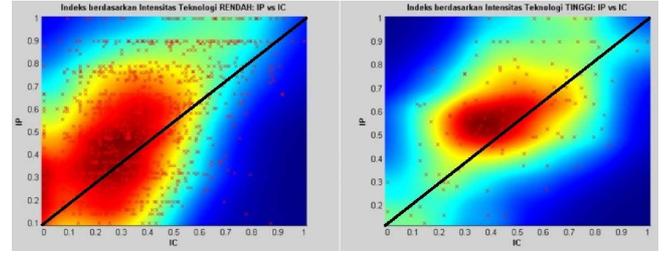}
\caption{The distribution of relationship between structural capital index and innovation performance index based on technology intensity (left: low-technology, right: high-technology).}
\label{fig:intensitas}
\end{figure}

The distribution of IC versus IP indexes are given in Figs. \ref{fig:intensitas}-\ref{fig:sdm}. The figures show the distribution of different correlations in each class under consideration. Fig. \ref{fig:intensitas} describes the distribution of correlation between structural capital and innovation performance indexes for low-technology (left figure) and high-technology (right figure) intensities. In Fig. \ref{fig:modal}, the distribution for small (left figure), medium (middle figure) and large (right figure) capitals are shown. Lastly, Fig. \ref{fig:sdm} shows the distribution for small (left figure) and large (right figure) number of employees. In all figures, the vertical axes denote the IP, while the horizontal ones are for IC. The more densed distribution the color is getting red and vice versa. Through the figures, one can 'qualitatively' get the insight of the acquired data rather than having average values without significant differences among them. 

\section{Summary and conclusion}

\begin{figure}[b!]
 \centering
 \includegraphics[width=85mm]{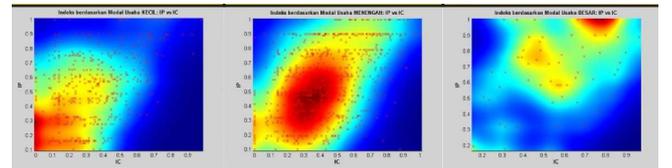}
\caption{The distribution of relationship between structural capital index and innovation performance index based on capital scale (left: small-capital, middle: medium-capital, right: large-capital).}
\label{fig:modal}
\end{figure}

The correlations between intellectual capital and innovation performance in the Indonesian manufacture industry have been presented. The study has been focused on the single component of structural capital. Though, the correlation is quite significant among different classes of companies based on its technology intensities, capital scales and also number of employees. Rather than representing the results as the standard statistical values, all of them have been presented in a more intuitive way using image manipulation based on the same obtained indexes. 

\begin{figure}[t]
 \centering
 \includegraphics[width=85mm]{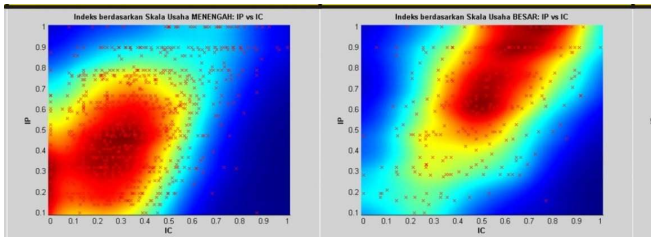}
\caption{The distribution of correlation between structural capital index and innovation performance index based on business scale (left: medium-scale, right: large-scale).}
\label{fig:sdm}
\end{figure}

From Fig 1, one can observe that mutual correlation appears only in companies with low technology intensity. It can be argued that this phenomena due to the fact that most of companies with high technology intensity developed in short time due to big investors rather than the natural evolution by enterpreuners in the industry. On the other hand, Figs. \ref{fig:modal} and \ref{fig:sdm} show that there are positive correlations among IP and IC. This means the innovation performance of most Indonesian manufactures is determined by its scale. 

More complete study should be performed to incorporate the whole components of intellectual capital including the human and relational capitals. The work in this topic is still in progress and will be published elsewhere. It is also interesting to compare the sequential data obtained from the recent survey in 2011 \cite{grace}.

\section*{Acknowledgment}

This work was supported by PKPP, the Indonesia Ministry of Research and Technology in FY 2011. LTH gratefully appreciate warm hospitality during this work at the Pappiptek LIPI. 

\bibliographystyle{IEEEtran}
\bibliography{ICIMTR2012}

\begin{thebibliography}{10}
\providecommand{\url}[1]{#1}
\csname url@rmstyle\endcsname
\providecommand{\newblock}{\relax}
\providecommand{\bibinfo}[2]{#2}
\providecommand\BIBentrySTDinterwordspacing{\spaceskip=0pt\relax}
\providecommand\BIBentryALTinterwordstretchfactor{4}
\providecommand\BIBentryALTinterwordspacing{\spaceskip=\fontdimen2\font plus
\BIBentryALTinterwordstretchfactor\fontdimen3\font minus
  \fontdimen4\font\relax}
\providecommand\BIBforeignlanguage[2]{{%
\expandafter\ifx\csname l@#1\endcsname\relax
\typeout{** WARNING: IEEEtran.bst: No hyphenation pattern has been}%
\typeout{** loaded for the language `#1'. Using the pattern for}%
\typeout{** the default language instead.}%
\else
\language=\csname l@#1\endcsname
\fi
#2}}

\bibitem{acs}
Z.~J. Acs and D.~B. Audretsch, Eds., \emph{Innovation and Technological Change:
  An International Comparison}.\hskip 1em plus 0.5em minus 0.4em\relax
  University of Michigan Press, 1991.

\bibitem{castro}
G.~M.~D. Castro, M.~D. Verde, P.~L. Saez, and J.~E.~N. Lopez,
  \emph{Technological Innovation: An Intellectual Capital-Based View}.\hskip
  1em plus 0.5em minus 0.4em\relax Palgrave Macmillan, 2010.

\bibitem{edquist}
C.~Edquist, Ed., \emph{Systems of Innovation: Technologies, Institutions and
  Organization}.\hskip 1em plus 0.5em minus 0.4em\relax Routledge, 1999.

\bibitem{grif}
M.~D. Griffiths, L.~Gundry, J.~Kickul, and A.~M. Fernandez, ``Innovation
  ecology as a precursor to entrepreneurial growth: A cross-country empirical
  investigation,'' \emph{Journal of Small Business and Enterprise Development},
  vol.~16, no.~3, pp. 375--390, 2009.

\bibitem{myt}
L.~Mytelka, ``Local systems of innovation in a globalized world economy,''
  \emph{Industry and Innovation}, vol.~7, no.~1, pp. 15--32, 2000.

\bibitem{solow}
R.~M. Solow, ``A contribution to the theory of economic growth,''
  \emph{Quarterly Journal of Economics}, vol.~70, no.~1, pp. 65--94, 1956.

\bibitem{oecd1}
\BIBentryALTinterwordspacing
OECD, ``National innovation systems,'' OECD Publications, 1997. [Online].
  Available: \url{http://www.oecd.org/dataoecd/35/56/2101733.pdf}
\BIBentrySTDinterwordspacing

\bibitem{oecd2}
\BIBentryALTinterwordspacing
------, ``Oslo manual: Guidelines for collecting and interpreting innovation
  data,'' OECD Publications, 2005. [Online]. Available:
  \url{http://www.oecd.org/dataoecd/35/56/2101733.pdf}
\BIBentrySTDinterwordspacing

\bibitem{leif}
L.~Edvinsson and M.~S. Malone, \emph{Intellectual Capital: Realizing your
  Company's True Value by Finding Its Hidden Roots}.\hskip 1em plus 0.5em minus
  0.4em\relax Harper Business, 1997.

\bibitem{maddock}
J.~Maddocks and M.~Beaney, \emph{Knowledge Management}, 2002.

\bibitem{mei}
{S. Meiningsih et.al.}, \emph{Research and Development in Indonesian
  Manufacture Industry}.\hskip 1em plus 0.5em minus 0.4em\relax LIPI Press,
  2009.

\bibitem{rini}
R.~Wijayanti and L.~T. Handoko, ``Generic index for visualizing the binary
  data,'' under submission.

\bibitem{grace}
{N. G. Berliana et.al.}, ``Research and development in indonesian manufacture
  industry,'' Pappiptek LIPI, Tech. Rep., 2011.

\end{thebibliography}

\begin{biography}[{\includegraphics[width=1in,height=1.25in,clip,keepaspectratio]{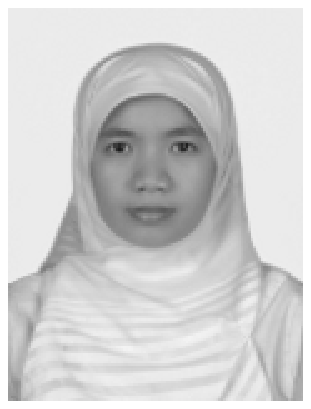}}]{Rini Wijayanti}
was born in Probolinggo, Indonesia on September 3, 1979. She graduated from Faculty of Science, University of Indonesia and Surabaya Institute of Technology. She has been involved in researches on data-mining, information system  and machine learning.
\end{biography}

\begin{biography}[{\includegraphics[width=1in,height=1.25in,clip,keepaspectratio]{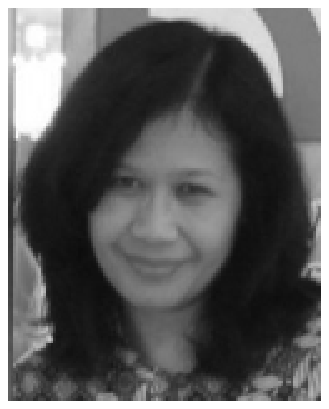}}]{N.G. Berliana}
 was born in Jakarta, Indonesia on September 24, 1966. She graduated from  Department of Mathematics and Science and Technology Study at University of Indonesia. She is now the Head of  Innovation Management and Science and Technology Indicator Division in at Pappiptek LIPI.
\end{biography}

\begin{biography}[{\includegraphics[width=1in,height=1.25in,clip,keepaspectratio]{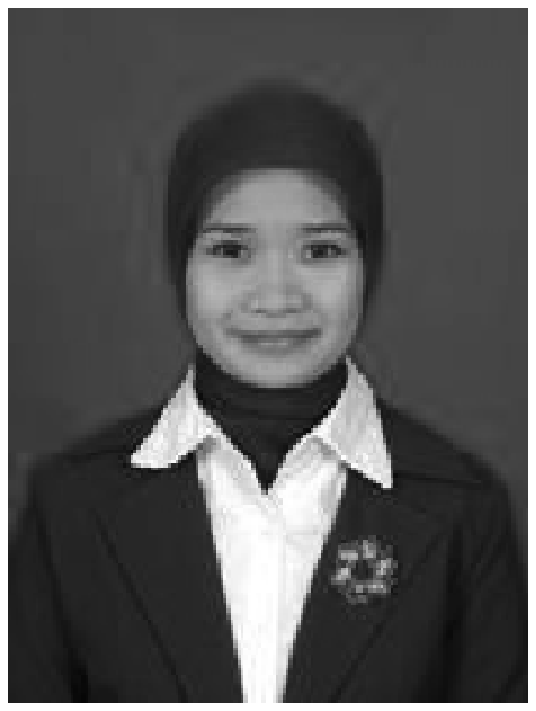}}]{I.M. Nadhiroh}
 was born in Padang, Indonesia on October 3, 1986. She is majoring statistics and graduated from  Bogor Institute of Agricultural. She is a young researcher at Pappiptek LIPI.
\end{biography}

\begin{biography}[{\includegraphics[width=1in,height=1.25in,clip,keepaspectratio]{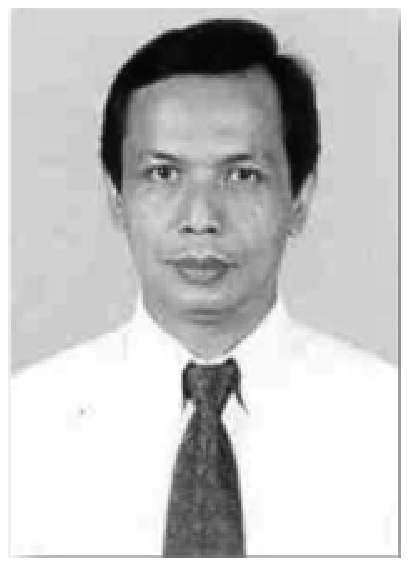}}]{E. Aminullah}
 was born in Bukittingi, Indonesia on October 5, 1955. He was a Research Professor of Technology Policy at  Pappiptek LIPI since 2006. He graduated from  Saitama University and University of Gajah Mada. His research is focused on Technology Policy and Economic Development.
\end{biography}

\begin{biography}[{\includegraphics[width=1in,height=1.25in,clip,keepaspectratio]{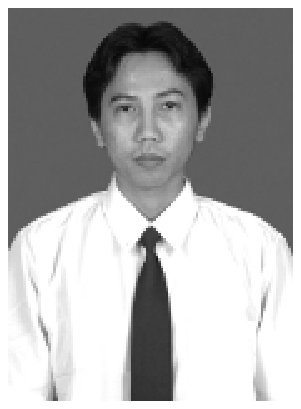}}]{Kusnandar}
 was born in Garut, Indonesia on July 23, 1979. He graduated from  Industry Engineering, Padjajaran University. He is a young researcher at Pappiptek LIPI.
\end{biography}

\begin{biography}[{\includegraphics[width=1in,height=1.25in,clip,keepaspectratio]{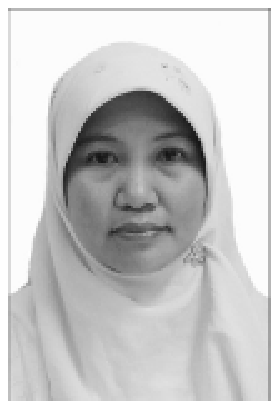}}]{T. Fizzanty}
 was born in Padang on March 28, 1968. She graduated from  University of  Queensland and Bogor Institute of Agricultural. She is the Head of Science and Technology Management System at Pappiptek LIPI.
\end{biography}

\begin{biography}[{\includegraphics[width=1in,height=1.25in,clip,keepaspectratio]{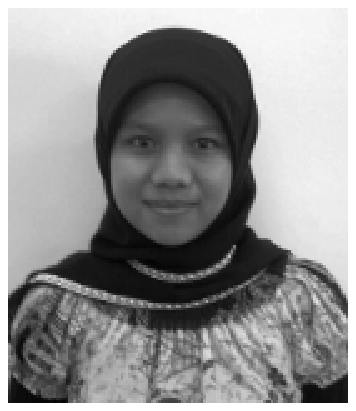}}]{N. Laili}
 was born in Surakarta, Indonesia on April 6, 1985. She graduated from  Industry Engineering, Sebelas Maret Surakarta University. She is a young researcher at Pappiptek LIPI.
\end{biography}

\begin{biography}[{\includegraphics[width=1in,height=1.25in,clip,keepaspectratio]{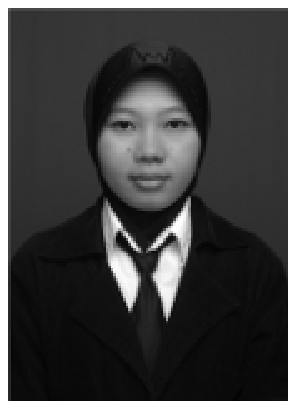}}]{R. Rahmaida}
 was born in Kebumen, Indonesia on October 12, 1986. She graduated from Department of Mathematics, Jenderal Soedirman University. She is a young researcher at Pappiptek LIPI.
\end{biography}

\begin{biography}[{\includegraphics[width=1in,height=1.25in,clip,keepaspectratio]{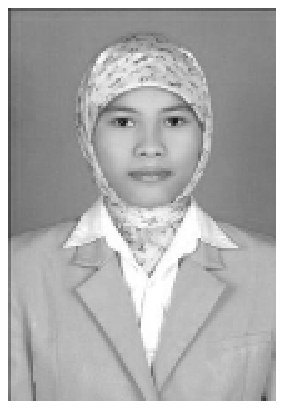}}]{S. Handayani}
 was born in Aceh, Indonesia on February 22, 1985. She graduated from Computer Science, University of Gajah Mada. She is a young researcher at Pappiptek LIPI.
\end{biography}

\begin{biography}[{\includegraphics[width=1in,height=1.25in,clip,keepaspectratio]{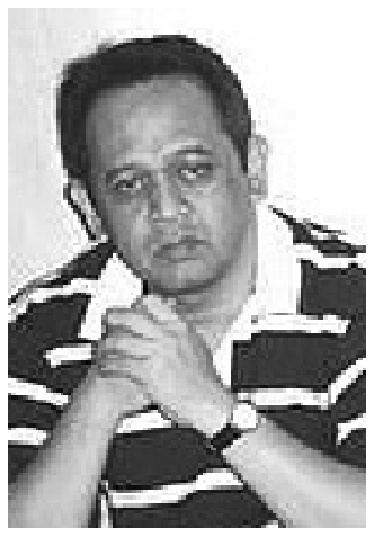}}]{Kusbiantono}
  was born in Banyuwangi, Indonesia on May 10, 1961. He graduated from Science and Technology Management, University of Indonesia. His research is focused on Research and Development Management and Innovation. \end{biography}

\begin{biography}[{\includegraphics[width=1in,height=1.25in,clip,keepaspectratio]{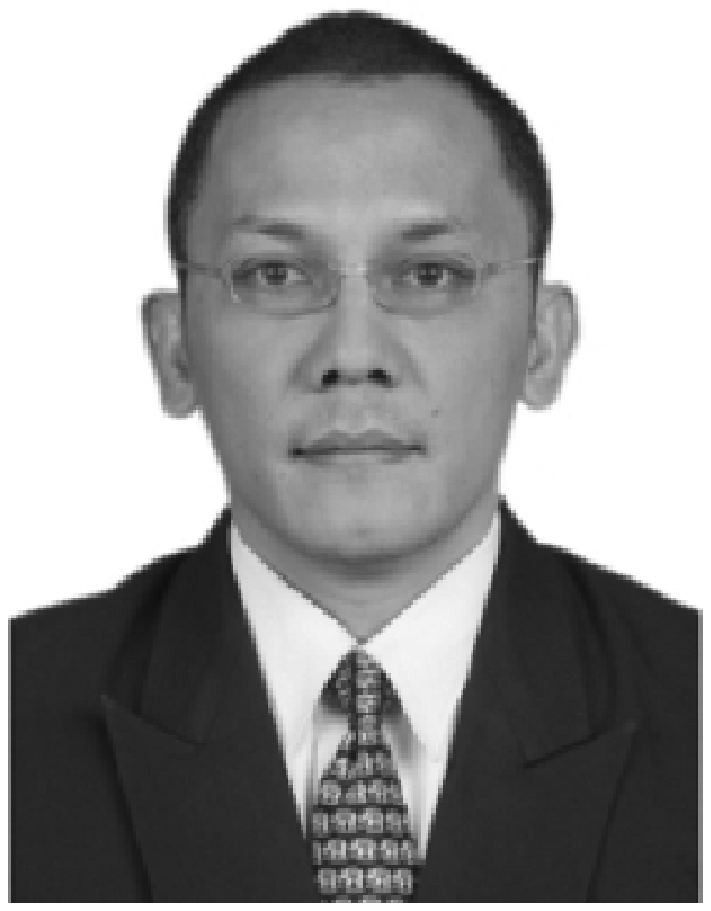}}]{L.T. Handoko}
 was born in Malang, Indonesia on May 7, 1968. He graduated from Theoretical Physics, Kumamoto University and Hiroshima University. His research area ranges from theoretical physics, data mining and computational science. He is a member of  IEEE since 2008.
\end{biography}

\end{document}